\documentclass[review,numbers,sort,compress]{elsarticle}

\usepackage{lineno}
\usepackage{graphicx}
\usepackage{latexsym}%
\usepackage[table]{xcolor}
\usepackage{amsfonts}
\usepackage{booktabs}
\usepackage{lmodern}
\usepackage{lineno}
\usepackage{soul}
\usepackage{enumitem}
\usepackage{comment}
\usepackage{relsize}
\usepackage{lipsum}
\usepackage{color}
\usepackage{siunitx}
\usepackage[cal=boondoxo]{mathalfa}
\usepackage[breaklinks=true]{hyperref}
\usepackage{breakcites}
\usepackage{blindtext, rotating}
\usepackage{framed} 
\usepackage{multicol} 
\usepackage{pifont}
\usepackage[parfill]{parskip}
\usepackage{nomencl}
\usepackage{colortbl}
\usepackage{tabularx}
\usepackage[defaultlines=3,all]{nowidow}
\usepackage{amssymb}
\usepackage{amsmath}
\usepackage{epstopdf}
\usepackage{setspace}
\usepackage{xspace}
\usepackage{subcaption}
\usepackage{epsfig}
\usepackage{placeins}
\usepackage{epstopdf}
\usepackage{wrapfig}
\usepackage{algorithm}
\usepackage[noend]{algpseudocode}
\usepackage{caption}
\usepackage[version=3]{mhchem}
\usepackage{lineno,hyperref}
\usepackage{cleveref}
\usepackage[font=small,skip=0pt]{caption}
\usepackage{geometry}
\usepackage{todonotes}
\usepackage[normalem]{ulem} 

\modulolinenumbers[5]
\definecolor{lavender}{rgb}{0.75, 0.58, 0.89}

\newcolumntype{M}[1]{>{\centering\arraybackslash}p{#1}}
\newcolumntype{P}[1]{>{\raggedright\arraybackslash}p{#1}}

\newcommand{\prt}{\partial}

\newcommand{\baa}{\mbox{\boldmath $a$}}

\newcommand{\bchi}{\mbox{\boldmath $\chi$}}

\newcommand{\bu}{\mbox{\boldmath $u$}}
\newcommand{\bw}{\mbox{\boldmath $w$}}
\newcommand{\by}{\mbox{\boldmath $y$}}
\newcommand{\bbf}{\mbox{\boldmath $f$}}
\newcommand{\bxi}{\mbox{\boldmath $\xi$}}
\newcommand{\bpsi}{\mbox{\boldmath $\psi$}}

\newcommand{\balpha}{\mbox{\boldmath $\alpha$}}
\newcommand{\lp}{\left(}
\newcommand{\rp}{\right)}

\setlist[itemize]{leftmargin=*}
\newif\ifcomments 
\commentsfalse
\commentstrue 

\definecolor{Green}{rgb}{0,0.5,0}

\definecolor{lightgray}{gray}{0.9}
\definecolor{amethyst}{rgb}{0.8, 0.0, 0.8}
\definecolor{aogreen}{rgb}{0.01, 0.75, 0.24}

\journal{Journal of \LaTeX\ Templates}

\begin{document}
\begin{frontmatter}
\title{Skeletal Reaction Models for Methane Combustion}

\author[mymainaddress]{Y.\ Liu}
\author[mymainaddress]{H.\ Babaee}
\author[mymainaddress]{P.\ Givi}
\author[mysecondaryaddress]{H.K.\ Chelliah}
\author[myteritaryaddress]{D.\ Livescu}
\author[mymainaddress]{A.G.\ Nouri\corref{mycorrespondingauthor}}
\address[mymainaddress]{Department of Mechanical Engineering and Materials Science, University of Pittsburgh, Pittsburgh, PA 15261, USA}
\address[mysecondaryaddress]{Department of Mechanical and Aerospace Engineering, University of Virginia, Charlottesville, VA 22904, USA}
\address[myteritaryaddress]{Los Alamos National Laboratory, Los Alamos, NM 87544, USA}
\cortext[mycorrespondingauthor]{Corresponding author \\ E-mail address: arash.nouri@pitt.edu.}
\begin{abstract}

A local-sensitivity-analysis technique is employed to generate new skeletal reaction models for methane combustion from the foundational fuel chemistry model (FFCM-1). The sensitivities of the thermo-chemical variables with respect to the reaction rates are computed via the forced-optimally time dependent (f-OTD) methodology. In this methodology, the large sensitivity matrix containing all local sensitivities is modeled as a product of two low-rank time-dependent matrices. The evolution equations of these matrices are derived from the governing equations of the system. The modeled sensitivities are computed for the auto-ignition  of methane at atmospheric and high pressures with different sets of initial temperatures, and equivalence ratios. These sensitivities are then analyzed to rank the most important (sensitive) species. A series of skeletal models with different number of species and levels of accuracy in reproducing the FFCM-1 results are suggested. The performances of the generated models are compared against FFCM-1 in predicting the ignition delay, the laminar flame speed, and the flame extinction. 
The results of this comparative assessment suggest the skeletal models with 24 and more species generate the FFCM-1 results with an excellent accuracy.

\end{abstract}
\end{frontmatter}

\section{Introduction}
\label{section:Intro} 

There is a continuing need to develop skeletal kinetics models for hydrocarbon combustion \cite{Kohse21,Gorban18,LL09}. These models are typically produced by systematic elimination of unimportant species and reactions from a detailed kinetics model, while maintaining its overall predictive ability~\cite{LL09,GM11,Smooke91,Peters1993}. 
Within the past several decades, a variety of  techniques have been proposed for this purpose, \textit{e.g.} local sensitivity analysis (LSA)~\cite{Turanyi90b,tomlin2013,vom2019,NBGCL22}, computational singular perturbation~\cite{CSP,neophytou2004reduced,lu2008criterion}, reaction flux analysis~\cite{Turanyi90a,wang:1991,SCGJ10}, and directed relation graph (DRG) and its variants~\cite{LL05(DRG),PP08(DRGEP),NS10(DRGEPSA)}. The LSA-based approaches, which are computationally costly for large kinetics models, contain techniques such as principal component analysis~\cite{BLK97,ECh11,PSDTS11,PS13,ME14,CIGP16,MICSP18} and species ranking construction~\cite{SFCFR16}. 
In LSA, the sensitivities can be computed either by solving a sensitivity equation (SE) or by solving an adjoint equation (AE). The latter can become quite costly for time-dependent sensitivity problems, since the AE must be solved in a forward-backward workflow requiring significant I/O costs for large chemical kinetic systems.

The f-OTD method is an on-the-fly reduced order modeling (ROM) technique, recently introduced for computing sensitivities in evolutionary dynamical systems \cite{DCB22}. Unlike, the traditional ROM techniques, the f-OTD does not require any offline data generation, and all the computations are carried out online. Reference \cite{NBGCL22} introduces a LSA-based skeletal kinetics reduction technique that benefits from the computational advantages of the f-OTD, and automatically eliminates unimportant reactions and species. In this approach, the sensitivity matrix \textit{i.e.} $S(t) \in \mathbb{R}^{n_{eq} \times n_r}$ is approximated by the product of two skinny matrices $U(t) = [\bu_1(t), \bu_2(t), \dots, \bu_r(t)] \in \mathbb{R}^{n_{eq} \times r}$, and $Y(t) = [\by_1(t), \by_2(t), \dots, \by_r(t)] \in \mathbb{R}^{n_r \times r}$ which contain the f-OTD modes and f-OTD coefficients, respectively; with $n_{eq}$ denoting the number of equations (or outputs), $n_r$ the number of independent parameters, $r \ll$ min\{$n_{eq}, n_r$\} the reduction size, and $S(t) \approx U(t)Y^T(t)$. As shown in \cite{PB20}, the f-OTD type decomposition is an equivalent decomposition to the dynamical low-rank approximation \cite{KL07}. The computed sensitivities are then analyzed locally to find and rank the most important (sensitive) species. Skeletal models are generated by selecting sufficient number of high ranked species in order to make accurate predictions of physical quantities of interest.

\begin{figure}[h!]
\centering 
 \includegraphics[width=16cm]{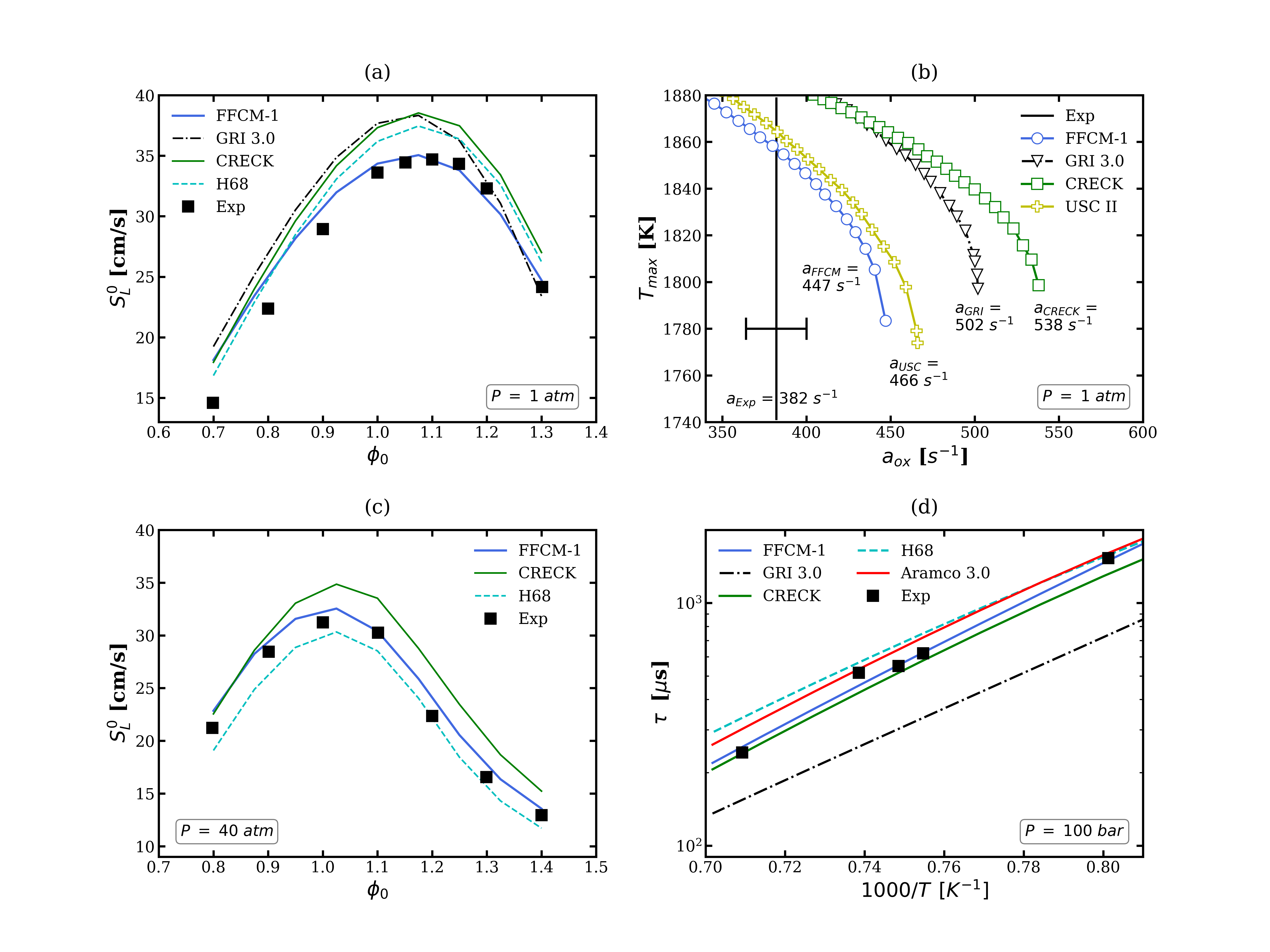}
 \caption{Validation of kinetic models against experimental data: (a) predictions of CH$_4$-air laminar flame speed vs. experimental data (square markers) of Lowry \textit{et al.}~\cite{LVK2011} at $\phi_0 = 1.0$ and T$_0$ = 298 $K$, (b) predictions of CH$_4$-air diffusion flame extinction vs. experimental data of Chelliah \textit{et al.}~\cite{Harsha12}, which is shown as a vertical line with 95\% confidence interval, (c) predictions of ``CH$_4$/17\% O$_2$/83\% He" flame speed vs experimental data (square markers) of Rozenchan \textit{et al.}~\cite{Rozenchan2002} at $\phi_0$ = $1.0$ and T$_0$ = 300 $K$, (d) ignition delay predictions benchmarked against experimental data (square markers) of Karimi \textit{et al.}~\cite{KO19} for a mixture of CH$_4$/O$_2$/Ar=3:6:91 (mole fraction).}
\label{FIG:exp_test}
\end{figure}

The objective of the present work is to generate accurate skeletal models for methane combustion at atmospheric and high pressure conditions. The f-OTD method is used for modeling local sensitivities of the temperature and the mass fractions with respect to reaction rates, and skeletal reduction is conducted by analyzing these sensitivities and ranking the most sensitive species.  Several detailed  kinetics models are available for methane combustion: GRI 3.0~\cite{GRI3}, USC Mech II~\cite{USC_MECH2}, and CRECK~\cite{CRECK} models are usually used for atmospheric pressures ($\sim$1 $atm$), while the foundational fuel chemistry model (FFCM-1)~\cite{FFCM-1}, AramcoMech model 3.0~\cite{ARAM3}, and a model of Hashemi \textit{et al.} (H68)~\cite{HP-methane} are utilized for high pressure conditions. Shock tube and/or rapid compression machine experiments are required at high levels of pressure to gain insight into the underlying kinetics~\cite{pierro22, shao20, ARAM2}. Here, the detailed kinetics model is chosen by benchmarking the predicted laminar flame speeds, the ignition delays, and the extinction curves against experimental data. Figures\ \ref{FIG:exp_test}(a) and \ref{FIG:exp_test}(c) portray the laminar premixed flame speed predictions by FFCM-1, GRI 3.0, CRECK, and H68. The experimental data at atmospheric and high pressures are taken from Lowry \textit{et al.}~\cite{LVK2011} and Rozenchan \textit{et al.}~\cite{Rozenchan2002}, respectively. It is shown that the flame speeds calculated by FFCM-1 and H68 are in good agreements with the experimental results. Figure\ \ref{FIG:exp_test}(b) shows that the FFCM-1 prediction for diffusion flame extinction is closer to the experimental data of Chelliah and co-workers~\cite{Harsha12} at 1 $atm$. Figure\ \ref{FIG:exp_test}(d) benchmarks the ignition delays as predicted by several detailed models against the experimental data of Karimi \textit{et al.}~\cite{KO19} at $100$ $bar$. Aramco 3.0, CRECK, and FFCM-1 indicate great performances in predicting ignition delays. Based on these comparisons, the FFCM-1  with 38 species and 569 irreversible reactions is selected as the starting bench-marked detailed kinetics model. 

\section{F-OTD for Skeletal Reduction}\label{section:skeletal_fOTD}

\subsection{Reduced-order modeling of the sensitivity with f-OTD}
\label{subsection:model_sens}

The temporal changes in mass fractions $\bpsi = [\psi_1, \psi_2, \dots, \psi_{n_s}]^T$ of chemical species and temperature $T$ in an adiabatic, constant pressure $P$, and spatially homogeneous chemical system of $n_s$ species reacting through $n_r$ irreversible reactions are governed by the following initial value problem~\cite{ZZT03,NBGCL22}:
\begin{equation}
\label{eq:baseODE}
\frac{d\xi_{i}}{dt} = f_{i} \lp \bxi,\balpha \rp, \ \ \bxi(0) = [\bpsi_0,T_0],
\end{equation}
where $\bxi = [\bpsi , T] \in \mathbb{R}^{n_{eq}}$,  $\bbf = [\bbf_{\bpsi} , f_T]$, $t$ is time, and $n_{eq} = n_s +1$.  In Eq.\ (\ref{eq:baseODE}), $\balpha = [1,1, \dots, 1] \in \mathbb{R}^{n_r}$ denotes the vector of sensitivity parameters perturbing the progress rate of reactions  $\mathcal{Q}_j = \alpha_j \mathcal{k}_{j} \prod_{m=1}^{n_s} (\rho \psi_m/W_m)^{\nu'_{mj}}$ with $j=1,2, \dots, n_r$. Here, $\rho$ denotes the density, $\nu'_{mj}$ is the molar stoichiometric coefficients of species $m$ in reaction $j$, and $W_m$ is the molecular weight of species $m$. A perturbation with respect to $\alpha_j$ around $\alpha_j=1$, implies  an infinitesimal perturbation of $\mathcal{Q}_j$. The parameter  $\mathcal{k}_{j}$ is the forward rate constant of reaction $j$, which is calculated via the modified Arrhenius model for elementary reactions~\cite{Williams1985b}. All reversible reactions are cast as irreversible reactions.  The matrix, $S(t)\in \mathbb{R}^{n_{eq} \times n_r}$, contains local sensitivity coefficients, $S_{ij}=\prt \xi_i / \prt \alpha_j$, and is evolved by solving the sensitivity equation:
\begin{equation}
 \label{eq:sensit_exact}
 \begin{split}
\frac{dS_{ij}}{dt} &=  \sum_{m=1}^{n_{eq}} \frac{\prt f_{i}}{\prt \xi_{m}} \frac{\prt \xi_{m}}{\prt \alpha_{j}} + \frac{\prt f_{i}}{\prt \alpha_{j}}  = \sum_{m=1}^{n_{eq}} L_{im} S_{mj} + F_{ij}, 
\end{split}
\end{equation}
\noindent where $L_{im} =  \frac{\prt f_{i}}{\prt \xi_{m}}$ and $F_{ij}=\frac{\prt f_{i}}{\prt \alpha_{j}}$ denote the Jacobian and the forcing matrices, respectively. 

In f-OTD,   the sensitivity matrix $S(t)$ is factorized into a time-dependent subspace in the $n_{eq}$-dimensional phase space of compositions represented by a set of f-OTD modes: $U(t) = [\bu_1(t), \bu_2(t), \dots, \bu_r(t)] \in \mathbb{R}^{n_{eq} \times r}$. These modes are orthonormal: $\bu_i^T(t) \bu_j(t) = \delta_{ij}$, where $\delta_{ij}$ denotes the  Kronecker delta function. The rank of $S(t)\in \mathbb{R}^{n_{eq} \times n_r}$ is $d=\mbox{min}\{n_{eq},n_r\}$, while the f-OTD modes represent a rank-$r$ subspace, where $r\ll d$. The sensitivity matrix is approximated via the f-OTD decomposition, $S(t) \approx U(t)Y^T(t)$,
where $Y(t) = [\by_1(t), \by_2(t), \dots, \by_r(t)] \in \mathbb{R}^{n_r \times r}$ is the f-OTD coefficient matrix. Therefore, each sensitivity coefficient $S_{ij}$ can be approximated as a finite sum of time-dependent terms: $S_{ij}(t)\approx \sum_{k=1}^r U_{ik}(t)Y_{jk}(t)$. The key characteristic of the model is that both $U(t)$ and $Y(t)$ are time-dependent, and evolve according to the closed form evolution equations extracted from the governing equations of the system
\cite{NBGCL22}:
\begin{subequations}
\label{eq:UY_evolution}
\begin{eqnarray} 
\label{eq:U_evolution}
\frac{dU}{dt} &=& QLU+QFYC^{-1}, \\
\frac{dY}{dt} &=& Y L_r^T +  F^T U,
\end{eqnarray}
\end{subequations}
\noindent where $L_r = U^T L U \in \mathbb{R}^{r \times r}$ is a reduced linearized operator, $Q= I-UU^{T}$ is the orthogonal projection onto the space spanned by the complement of $U$, and $C=Y^TY \in \mathbb{R}^{r \times r}$ is a \emph{correlation matrix}. The model constructed in this way is able to capture sudden transitions associated with the largest finite time Lyapunov exponents~\cite{BS16,BFHS17}. Equation (\ref{eq:UY_evolution}) is a coupled system of ODEs and constitutes the f-OTD evolution equations. The matrix $C(t)$ is, generally full, implying that the f-OTD coefficients are correlated. The f-OTD modes can become uncorrelated and energetically ranked by rotating the modes along the eigen-direction of the matrix $C$. When the energy of some of the f-OTD modes are dominant and their singular values are orders of magnitude larger than the energy associated with the other modes, it would be better to use the rank-adaptive and sparse-sampling f-OTD methodology \cite{donello2023cur, naderi2023adaptive}. This would circumvent numerical instabilities due to the presence of $C^{-1}$ in Eq.\ (3a). The f-OTD modes align themselves with the most instantaneously sensitive directions of the composition evolution equation when perturbed by $\balpha$. These directions can be unambiguously defined as the rank-$r$ reduction of instantaneous singular value decomposition (SVD) of $S(t)$. It has been shown that f-OTD closely approximates the instantaneous SVD of  $S(t)$, and also approximates sensitive directions without having access to any data on full-dimensional $S(t)$~\cite{NBGCL22}. Similar methodologies via time-dependent bases have also been utilized in stochastic ROM \cite{SL09a,CHZI13,Babaee:2017aa,B19,PB20}, flow control \cite{BMS18a},  rare events predictions \cite{FS16a}, and ROM of transport equations \cite{RNB21, amiri2022fly}. 

\subsection{Important reactions \& species}
\label{subsection:Reaction_selec_f-OTD}

In f-OTD, modeled sensitivities are computed in factorized format by solving Eqs.\ (\ref{eq:UY_evolution}), in addition to Eq.\ (\ref{eq:baseODE}), and the values of $\bxi$, $U$, and $Y$ are stored at resolved time steps $t_i \in [0,t_f]$. 
At each resolved time step, and for each case, the eigen decomposition of $S^T S\in \mathbb{R}^{n_r \times n_r}$ as $S^T S=A \Lambda A^T $ are computed along with $\bw \in \mathbb{R}^{n_r\times 1} = ( \Sigma \lambda_i |\baa_i|)/( \Sigma \lambda_i) \in \mathbb{R}^{n_r}$. The $\bw$ vectors are basically the average of eigenvectors of $S^T S$ matrix weighted based on their associated eigenvalues.  This prevents the f-OTD method from dealing with each eigenvector ($\baa_i$) separately. The first sorted eigenvalue ($\lambda_1$) is usually orders of magnitude larger than the others, which implies $\bw(t) \approx |\baa_1(t)|$. Each element of $\bw$, \textit{i.e.} $w_i$, is positive and associated with a certain reaction ($i$th reaction). The larger the $w_i$ value, the more important the reaction $i$ is. The variable $\chi_i$ is used to define as the highest value associated with $w_i$, \textit{i.e.} $\chi_i= max_t(w_i(t))$.

The elements of $\bchi$ vector are sorted in descending order to find the indices of the most important reactions in the detailed model. Species are also sorted based on their first presence in the sorted reactions, \textit{i.e.} the species which first shows up in a higher ranked reaction are more important than a species that first participate in a lower ranked reaction. This yields in a reaction and species ranking based on the $\bchi$ vector. Finally, a set of species are chosen by setting the threshold $\chi_{\epsilon}$ on the element of $\bchi$ vector and eliminating species whose associated $\chi_i$ values are smaller. The skeletal model reduction is reaction based, thus any non-reactive species with non-zero mass fraction in the initial condition should be manually added to the skeletal model, such as N$_2$ and/or Ar.

In summary, the f-OTD method instantaneously observes the ignition system, and sorts the reactions based on their effects on sensitivities to find the most important species. These species and the reactions which connect them together create the skeletal models.  In chemical kinetic systems, perturbations with respect to ``fast" reactions generate very large sensitivities for short time periods, but these sensitivities vanish as $t\rightarrow\infty$. On the other hand, perturbations with respect to ``slow" reactions generate smaller and more sustained sensitivities. Since the approach is based on the instantaneous observation of sensitivities, both slow and fast reactions can leave an imprint on the instantaneous normalized reaction vector ($\bw$). However, if the sensitivities associated with these reactions  would be averaged together over a period of time, then the smaller sensitivities associated with slow reactions could be out-weighted by the large sensitivities associated with fast reactions. 

\section{Skeletal Reduction of FFCM-1}
\label{comb_pressure}
The f-OTD method is employed to compute local sensitivities based on Eq.\ (\ref{eq:UY_evolution}) for atmospheric and high pressure ignition of methane. The $U$ and $Y$ matrices are initialized by solving the sensitivity equation (Eq.\ (\ref{eq:sensit_exact})) for few time steps, and then performing singular value decomposition on the sensitivity matrix. All f-OTD simulations are performed with $r=2$. The initial conditions for temperature, pressure, and equivalence ratio are $T_0 \in \{1200, 1500, 2000\}$ $K$, $P\in \{1, 20, 40, 60, 80, 100\}$ $atm$, and $\phi\in \{0.5, 1.0, 1.5\}$, respectively. Therefore, in total 48 cases are considered. 
As described in \S2, the outcome of the reduction process is a series of skeletal models as listed in Table\ \ref{table:models}. The generated f-OTD models are labeled by ``f-OTD-X" in which ``X" denotes the number of species included in the model. In Table\ \ref{table:models}, $n_r$ shows the number of irreversible reactions in each of the models. 

\begin{figure}[h!]
\centering 
 \includegraphics[width=10cm]{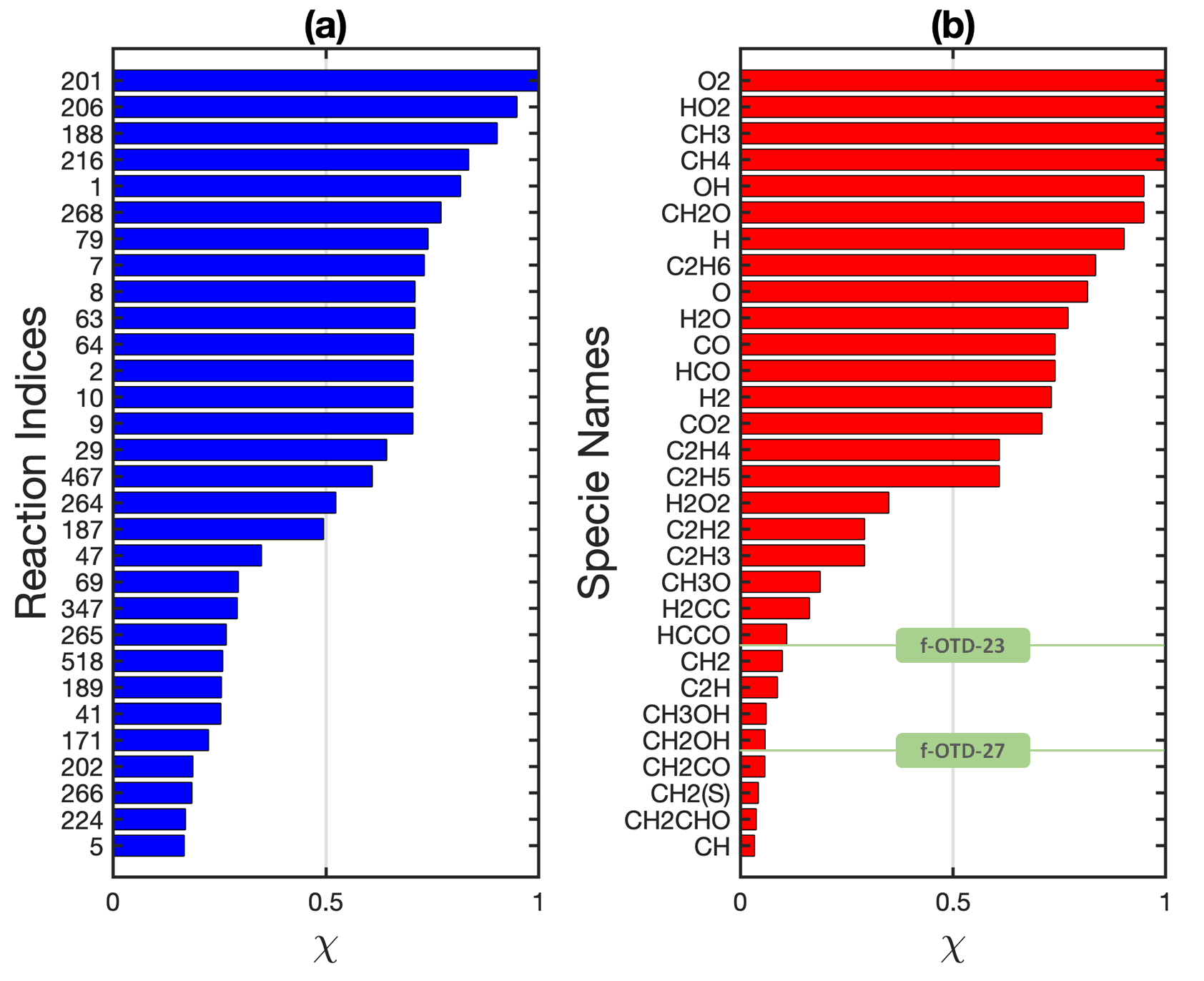}
 \caption{The FFCM-1 skeletal reduction for methane-air: sorted (a) reactions and (b) species based on their importance as determined by their associated $\chi$ values.}
\label{FIG:chi}
\end{figure}

\newcolumntype{?}{!{\vrule width 1.5pt}}
\begin{table}
\caption{FFCM-1 based skeletal models}
\label{table:models}
\scriptsize
\centering
\begin{tabular}{?M{2.0cm}?M{1.8cm}?M{1.8cm}?M{1.8cm}?}
\specialrule{1.5pt}{0pt}{0pt}
Model &  $n_s$ & $n_r$ & $\chi_{\epsilon}$ \\ 
\specialrule{1.5pt}{0pt}{0pt}
FFCM-1  & 38  & 569 & 0 \\ \hline
f-OTD-27 & 27 & 319 & 0.0583 \\ \hline
f-OTD-25 & 25  & 261 & 0.0877\\ \hline
f-OTD-24 & 24  & 247 & 0.0992\\ \hline
f-OTD-23 & 23  & 213 & 0.1093\\
\specialrule{1.5pt}{0pt}{0pt}
\end{tabular}
\end{table}

\subsection{Reactions and species ranking}
Figure \ref{FIG:chi} shows the species and reaction rankings based on the process described in \S\ref{subsection:Reaction_selec_f-OTD}. Reactions R201 (O$_2$ + CH$_4$ $\rightarrow$ CH$_3$ + HO$_2$), R206 ( CH$_3$ + O$_2$ $\rightarrow$ OH + CH$_2$O), and R188 (CH$_4$ (+M) $\rightarrow$ CH$_3$ + H (+M)) are identified as the top three most important ones in FFCM-1 for methane ignition. The oxidation of methane is initiated by its reaction with molecular oxygen (R201). The methyl radical (CH$_3$) then reacts with the molecular oxygen to form OH + CH$_2$O via R206. As for the species, O$_2$, HO$_2$, CH$_3$, CH$_4$, OH, CH$_2$O, H, C$_2$H$_6$, O, H$_2$O are the top 10 most important ones. The f-OTD-23 model contains the top 22 ranked species in Fig.\ \ref{FIG:chi}(b) plus nitrogen (non-reactive species with non-zero initial mass fractions). The f-OTD-24, f-OTD-25, f-OTD-26, and f-OTD-27 models are produced by adding, respectively, CH$_2$, C$_2$H, CH$_3$OH, and CH$_2$OH and their reactions step by step to f-OTD-23.

\subsection{Skeletal model validation}
The performances of the reduced models are assessed by comparing their estimations for the ignition delays, the laminar flame speeds, and the counter-flow extinction strains against FFCM-1 for different mixtures. Here, the ignition delay is defined as the time required by carbon monoxide (CO) to reach its maximum production rate. The flame speeds and the extinction curves are generated by Cantera~\cite{Cantera}  with a plug flow boundary condition assumption~\cite{fiala2014} for extinction simulations. 

\begin{figure}[h!]
\centering 
 \includegraphics[width=15cm]{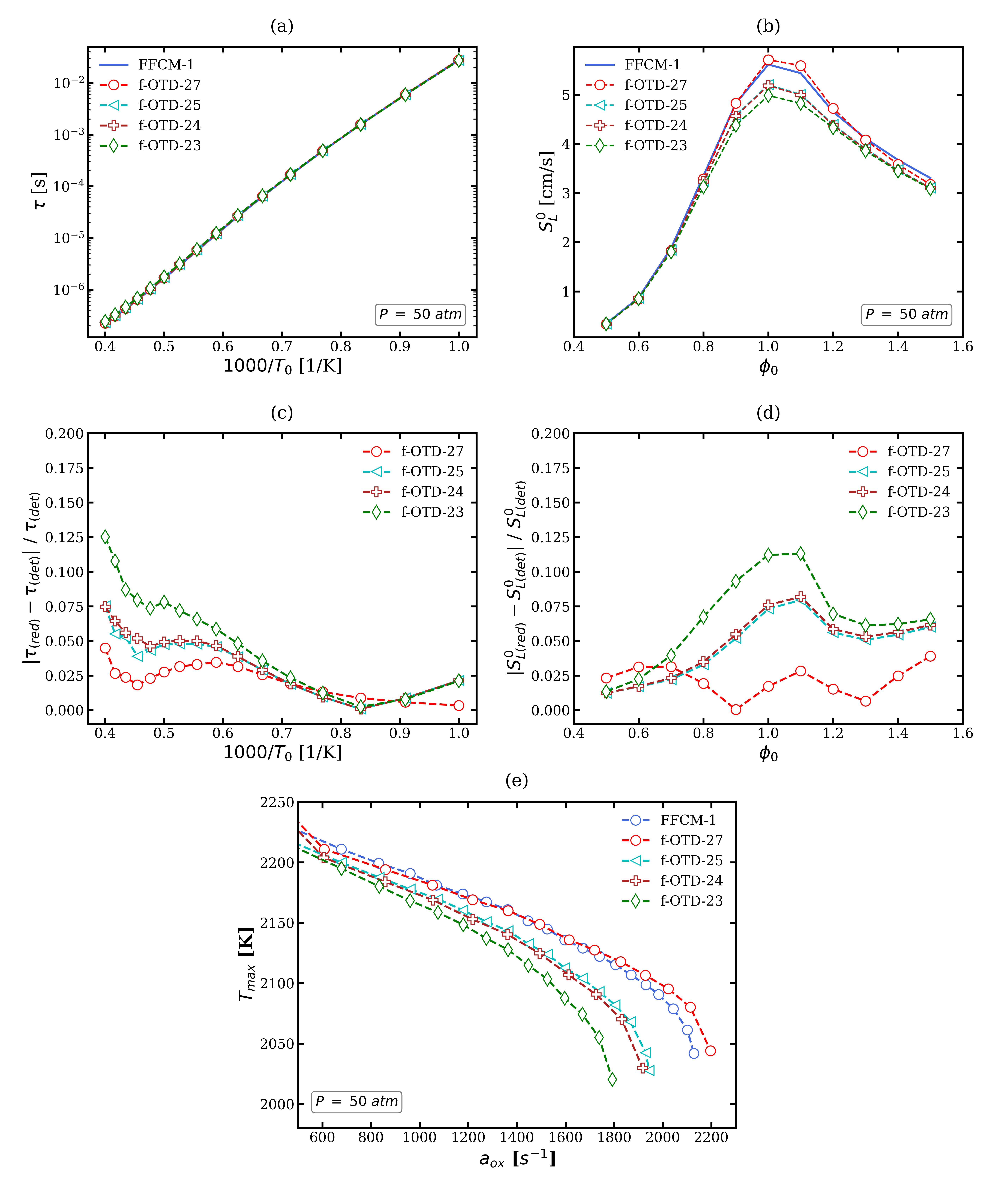}
 \caption{Predictions via f-OTD skeletal models at $P=50$ $atm$: (a) ignition delays, $\phi_0 = 1.0$, (b) flame speeds, $T_0 = 300$ $K$, (c) relative errors of skeletal models in terms of the ignition delays, (d) relative errors of skeletal models in terms of the flame speeds, (e) nonpremixed flame extinction with $T_0 = 300$ $K$.}
\label{FIG:IDFS50}
\end{figure}

\begin{figure}[h!]
\centering 
 \includegraphics[width=15cm]{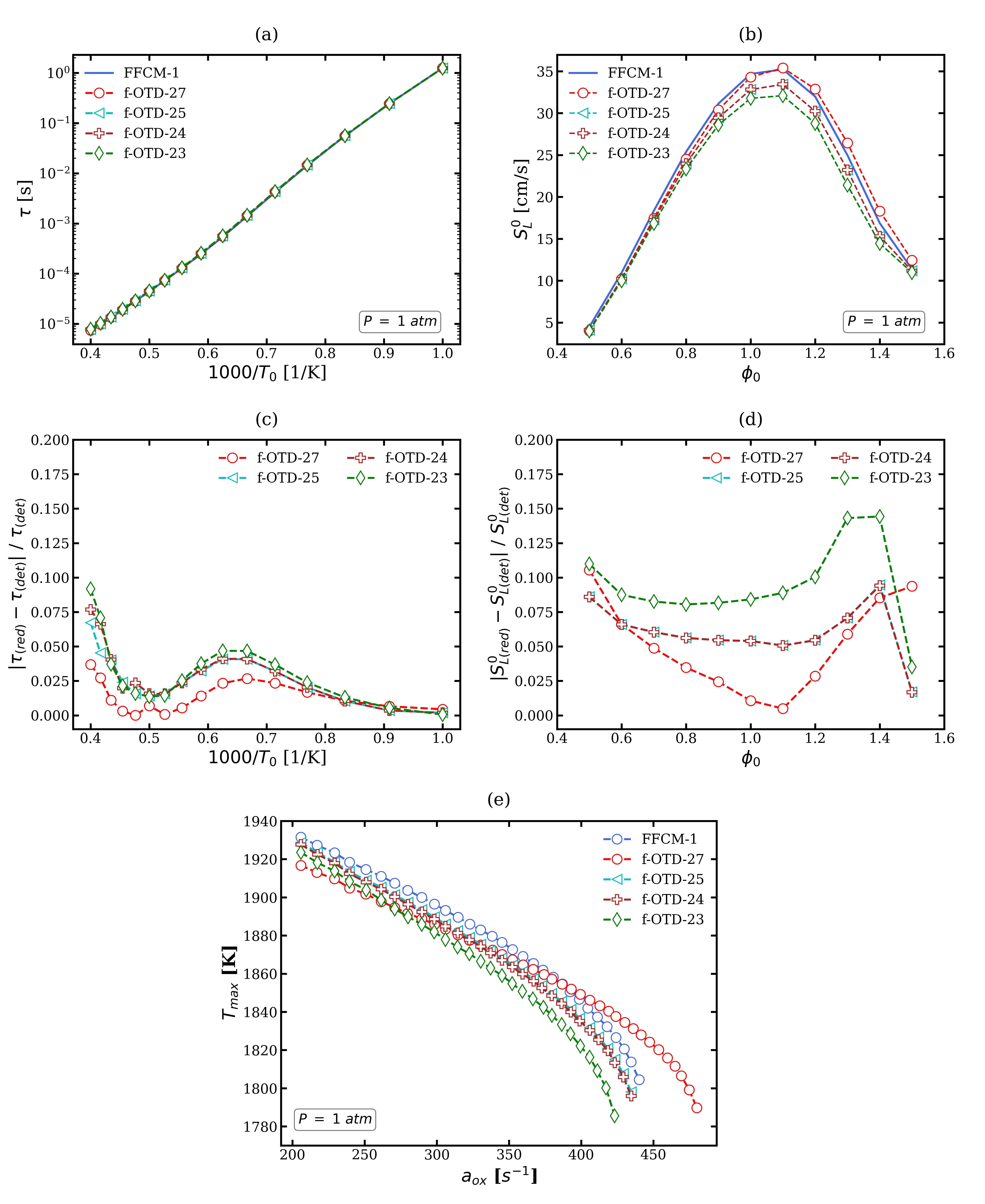}
 \caption{Predictions via f-OTD skeletal models at $P=1$ $atm$: (a) the ignition delays, $\phi_0 = 1.0$, (b) the flame speeds, $T_0 = 300$ $K$, (c) relative errors of skeletal models in terms of ignition delays, (d) relative errors of skeletal models in terms of the flame speeds, (e) nonpremixed flame extinction, $T_0 = 300$ $K$.}
\label{FIG:IDFS1}
\end{figure}

Figure\ \ref{FIG:IDFS50} compares the f-OTD results at $P=50$ $atm$ against FFCM-1. Figure\ \ref{FIG:IDFS50}(a) and (c) show that f-OTD models with $n_s \geq 24$ predict  ignition delays with less than 10\%  error for $T_0 \in [1000,2500] K$, and less than 5\%  error for $T_0 \in [1000,1670] K$. The f-OTD-27 is the most accurate skeletal model for estimating ignition delays at $P=50$ $atm$ with less than 5\% overall error. Figure\ \ref{FIG:IDFS50}(b) and (d) portray the ability of f-OTD-24, f-OTD-25, and f-OTD-27 models in reproducing the laminar flame speeds of FFCM-1 with less than 10\% error. The maximum error of f-OTD-27 for flame speed estimations is 3\%. Figure\ \ref{FIG:IDFS50}(e) shows that the f-OTD model with 27 species reproduces the extinction curve of FFCM-1 almost exactly. Altogether, the accuracy of f-OTD models in regenerating FFCM-1 results improves with increasing the number of species. Moreover, f-OTD-24 and f-OTD-25 show very similar predictions. This means that while C$_2$H is an important species, including it in f-OTD-25 does not change the predictions significantly.

Figure\ \ref{FIG:IDFS1} shows the performance of f-OTD models at $P=1$ $atm$. Figures\ \ref{FIG:IDFS1}(a), (b), (c), and (d) indicate better performances of the models in predicting the ignition delays and the laminar flame speeds by increasing $n_s$. The f-OTD-27 is the most accurate model with less than 5\% error in estimating ignition delays, and 10\% error in predicting laminar flame speeds. However, the flame speeds calculated by f-OTD-27 have larger errors for very lean and very rich fuels, \textit{i.e.} $\phi_0<0.6$ and $\phi_0>1.4$. 
Altogether, Fig.\ \ref{FIG:IDFS1} suggests that at atmospheric pressure, f-OTD model with $n_s\ge24$ provides reasonable predictions for the three quantities as considered.

\begin{figure}[h!]
\centering 
 \includegraphics[width=8cm]{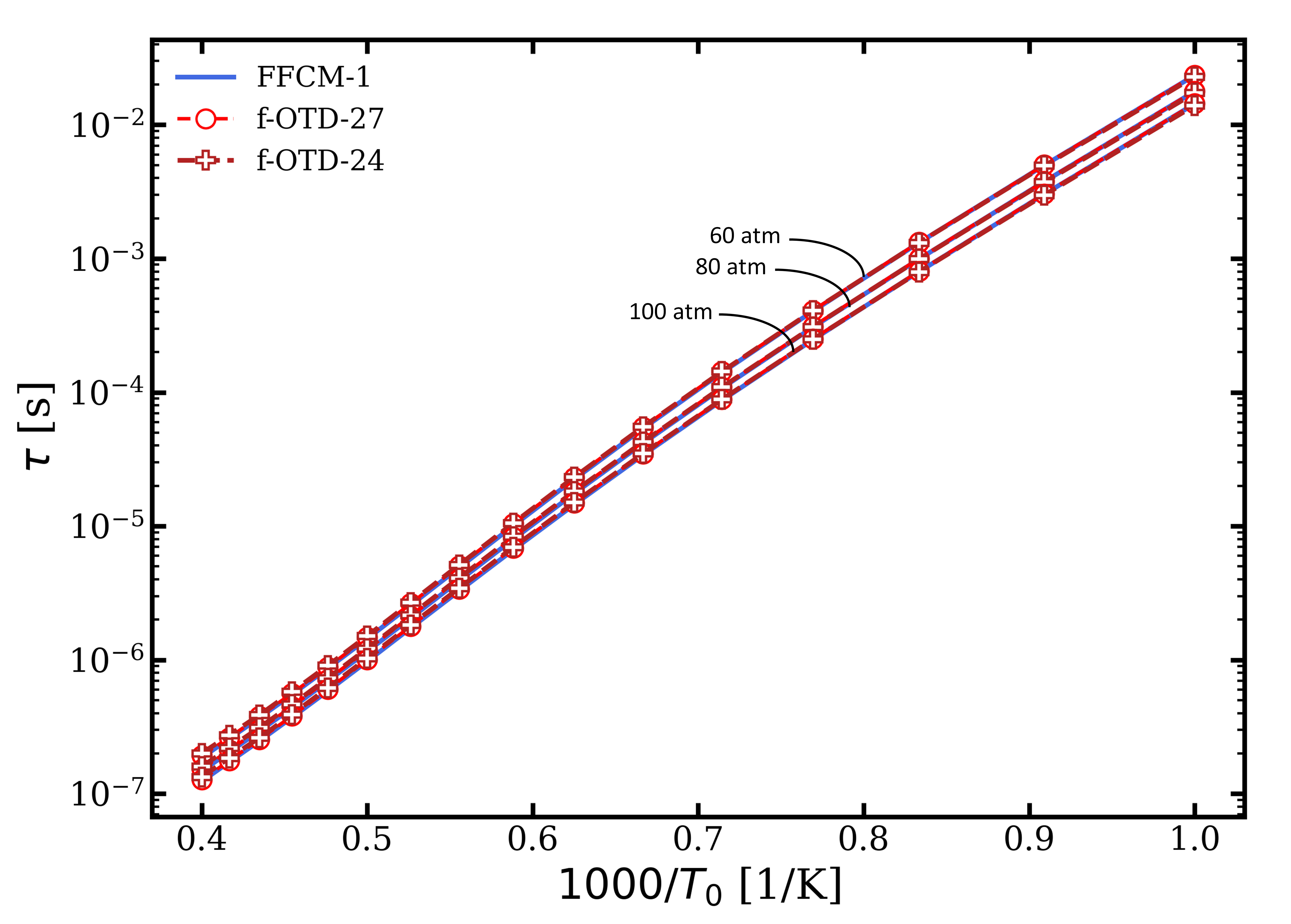}
 \caption{Predictions via f-OTD-24 and f-OTD-27: Ignition delays for $P=60,80$, and $100$ $atm$ with $\phi_0=1.0$.}   
\label{FIG:id_p_vali}
\end{figure}

\begin{figure}[h!]
\centering 
 \includegraphics[width=16cm]{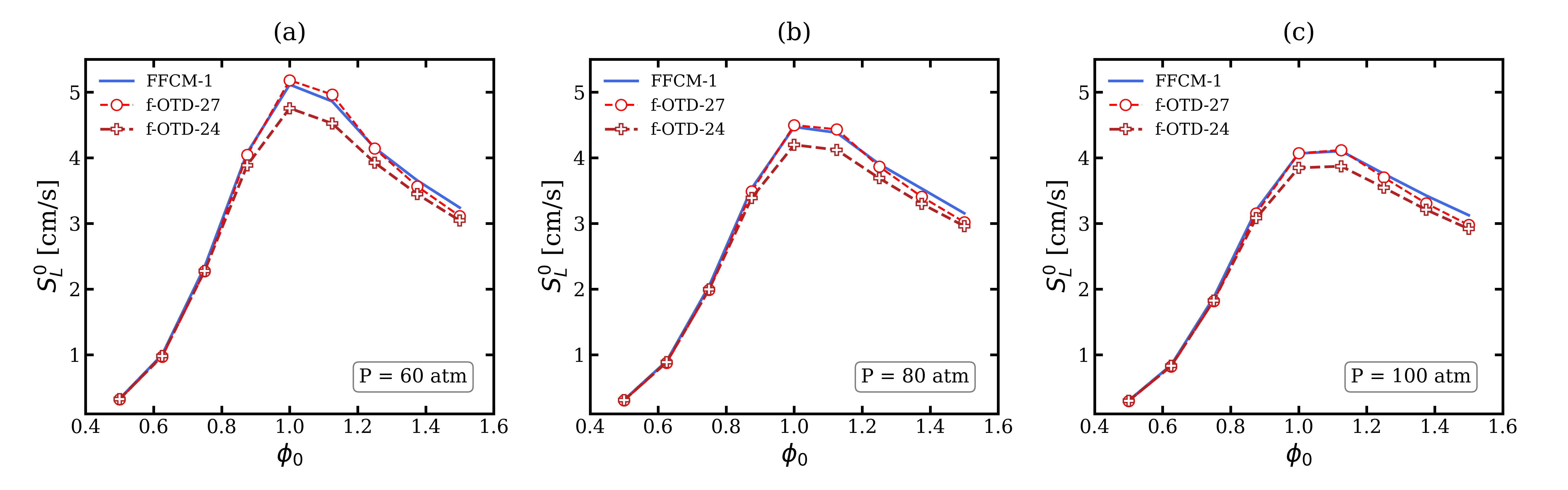}
 \caption{Predictions via f-OTD-24 and f-OTD-27: Laminar flame speeds for $P=60,80$, and $100$ $atm$ with $T_0=300$ $K$. }   
\label{FIG:fs_p_vali}
\end{figure}

\begin{figure}[h!]
\centering 
 \includegraphics[width=16cm]{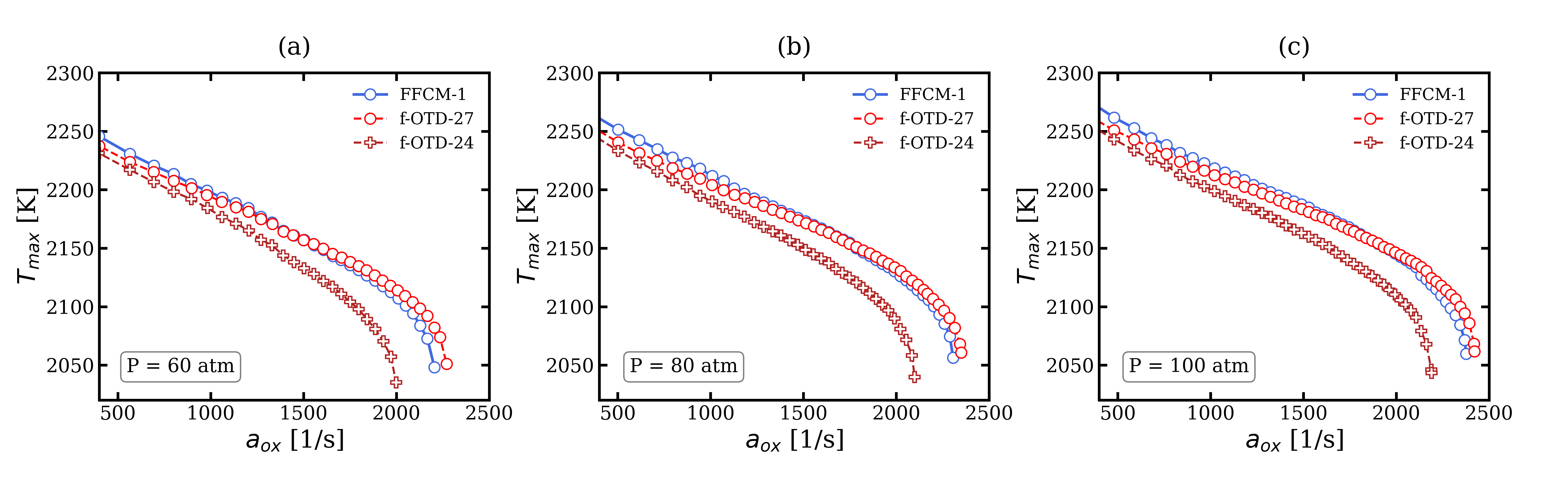}
 \caption{Predictions via f-OTD-24 and f-OTD-27: Diffusion flame extinction for $P=60,80$, and $100$ $atm$ with $T_0=300$ $K$.}   
\label{FIG:fe_p_vali}
\end{figure}

\begin{figure}[h!]
\centering 
 \includegraphics[width=16cm]{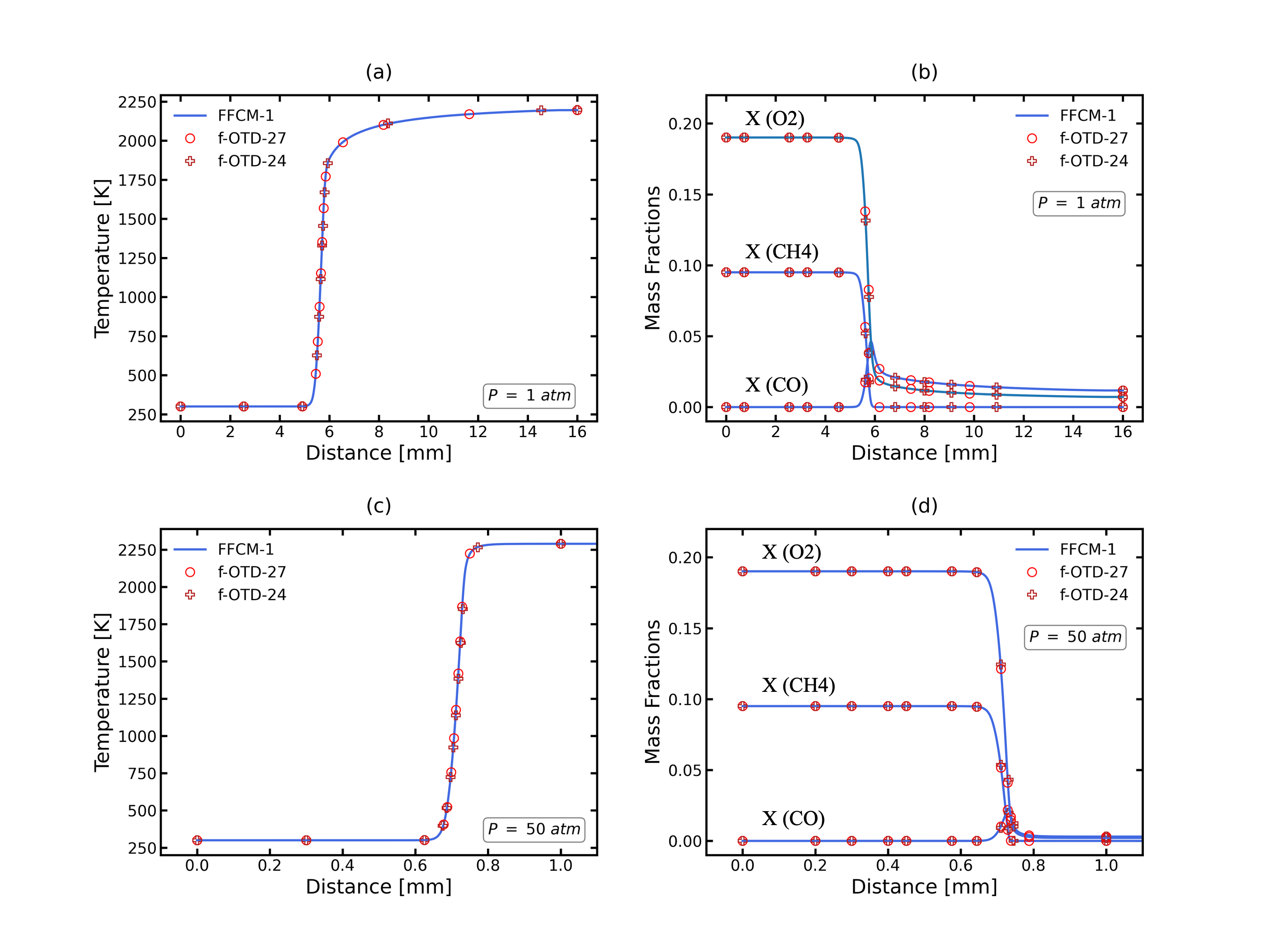}
 \caption{Prediction via f-OTD-24 and f-OTD-27: 1 D freely-propagating, premixed flame structure with $T_0$ $=$ $300$ $K$, $\phi_0$ $=$ $1.0$, and $P=1\&50$ $atm$}
\label{FIG:flame_structure}
\end{figure}

Figures\ \ref{FIG:id_p_vali} - \ref{FIG:fe_p_vali} demonstrate the  ability of f-OTD-24 and f-OTD-27 in reproducing the FFCM-1 results, over a wide range of pressures ($P=60$ $atm$, $80$ $atm$, and $100$ $atm$). Figure\ \ref{FIG:id_p_vali} compares the ignition delays predicted by these f-OTD models against FFCM-1, and Fig.\ \ref{FIG:fs_p_vali} indicates that f-OTD-24 and f-OTD-27 estimate the FFCM-1 flame speeds accurately. The f-OTD-27 predicts the laminar flame speeds reasonably well, with its worst predictions at $\phi_0>$1.2 where the relative error is still less than 10\%. Figure\ \ref{FIG:fs_p_vali}  also shows that the maximum flame speed decreases by increasing the pressure from $60$ $atm$ to $100$ $atm$. Figure\ \ref{FIG:fe_p_vali} compares the extinction curves of f-OTD-24 and f-OTD-27 against FFCM-1. Both models perform very well in this high pressure condition. Figure \ref{FIG:flame_structure} shows the structure of freely-propagating, premixed flame of methane at two different pressures, \textit{i.e.} $P=1$ $atm$ and $P=50$ $atm$. f-OTD-24 and f-OTD-27 reproduce the temperature and mass fractions predictions of FFCM-1 with a good accuracy.

\section{Conclusions}
\label{section:conclusion}
New skeletal kinetics models are generated from the FFCM-1 for the methane combustion at atmospheric and high pressure conditions. In the reduction process, local sensitivities are computed by the f-OTD method for the auto-ignition problem with different sets of initial conditions. The calculated sensitivities are then analyzed, and the 38 species of the FFCM-1 are ranked based on their importance. Different skeletal models with different levels of accuracy are generated by selecting more high ranked species. The ignition delays, the laminar flame speeds, and the diffusion flame extinctions predicted by the generated models are benchmarked against FFCM-1. The results show the model with 27 species and 319 irreversible reactions accurately reproduces the predictions of FFCM-1 at all conditions. The model with 24 species and 247 irreversible reactions also provides reasonable predictions. Therefore, f-OTD-24 and f-OTD-27 are the recommended skeletal models for the FFCM-1 over the observed range of pressures, temperatures, and equivalence ratios. The Cantera format (.cti) of these skeletal models are supplied in the Appendix. 

The f-OTD method has demonstrated its ability for reduced order modeling of local sensitivities in time dependent combustion systems, and is recommended for future multi-dimensional reacting flow simulations.  The described local sensitivity analysis technique for skeletal reduction is also recommended for  other detailed kinetics models. This technique does not require a priori expert knowledge of chemistry, thus can be used for skeletal reduction of very large reaction networks, \textit{e.g.} heavy hydrocarbon fuels like JP-10 (C$_{10}$H$_{16}$). For future work the f-OTD method is recommended for skeletal reduction of very large reaction networks, \textit{i.e.} with $\mathcal{O}(1000)$ species, by implementing rank-adaptive and sparse-sampling techniques ~\cite{donello2023cur, naderi2023adaptive}.

\section*{Acknowledgments}
This work has been co-authored by an employee of Triad National Security, LLC which operates Los Alamos National Laboratory under Contract No. 89233218CNA000001 with the U.S. Department of Energy/National Nuclear Security Administration. The work at Pitt is sponsored by the NSF under Grant CBET-2042918 and Grant CBET-2152803. Computational resources are provided by the Center for Research Computing (CRC) at Pitt.
\section*{Appendix}
All reduced mechanisms developed in this work can be accessed via this \href{https://github.com/Tom-Y-Liu/Chemical-Mechanisms.git}{\underline{link}}. 
\typeout{}
\bibliography{Methane_YL,ROM_combustion}

\end{document}